\documentclass[10pt,draft]{dis03}
\usepackage{epsf,amsmath}

\def\beq{\begin{equation}}
\def\eeq{\end{equation}}
\def\beqa{\begin{eqnarray}}
\def\eeqa{\end{eqnarray}}

\begin{document}

\title{A MASTER FORMULA FOR NNLO SOFT AND VIRTUAL\\
QCD CORRECTIONS
\thanks{The author's research has been supported 
by a Marie Curie Fellowship 
of the European Community programme ``Improving Human Research Potential'' 
under contract number HPMF-CT-2001-01221.}}

\author{NIKOLAOS KIDONAKIS \\
Cavendish Laboratory, University of Cambridge\\
Cambridge CB3 0HE, England\\
E-mail: kidonaki@hep.phy.cam.ac.uk }

\maketitle

\begin{abstract}
\noindent I present a master formula for the next-to-next-to-leading order
(NNLO) soft and virtual QCD corrections for any process
in hadron-hadron and lepton-hadron colliders. The formula is
derived from a unified threshold resummation formalism.
Applications to various QCD processes are discussed.
\end{abstract}

\section{Introduction} 
Calculations of total and differential cross sections in perturbative QCD
can be schematically represented by
\beq
\sigma=\sum_f \int \left[ \prod_i  dx_i \, \phi_{f/h_i}(x_i,\mu_F^2)\right]\,
{\hat \sigma}(s,t_i,\mu_F,\mu_R) 
\eeq
with $\sigma$ the physical cross section, ${\hat \sigma}$ the
perturbatively calculable hard scattering factor,
and $\phi_{f/h_i}$ the parton distribution for parton $f$ in hadron
$h_i$. 

The partonic hard-scattering factors ${\hat \sigma}$ include soft and virtual 
corrections, arising from soft-gluon emission and loop diagrams, which 
manifest themselves as plus distributions and delta functions with respect
to a kinematical variable $x_{th}$ that measures distance from threshold. 
In single-particle-inclusive (1PI) kinematics, $x_{th}$ is
often called $s_4=s+t+u-\sum m^2$, and the plus distributions are
${\cal D}_l(s_4)\equiv[\ln^l(s_4/M^2)/s_4]_+$, with $l \le 2n-1$ at $n$th 
order in $\alpha_s$ beyond the leading order.
In pair-inclusive (PIM) kinematics, the relevant distributions
are ${\cal D}_l(z) \equiv [\ln^l(1-z)/(1-z)]_+$
with $z=Q^2/s$ ($Q^2$ is of the pair). 
These distributions can be formally resummed to all orders
in $\alpha_s$ [1-5].
Here I present a unified approach and a master formula \cite{NKNNLO}
for explicitly calculating these corrections at NNLO for total and 
differential  cross sections for any process in

$\bullet$ hadron-hadron and lepton-hadron colliders \hspace{5mm} 
$\bullet$ 1PI and PIM kinematics

$\bullet$ simple and complex color flows \hspace{5mm}
$\bullet$ ${\overline{\rm MS}}$ and DIS factorization schemes

\section{NNLO master formula}
I begin by presenting a unified formula for the threshold resummed
cross section in moment space for an arbitrary process
\beqa
{\hat{\sigma}}^{res}(N) &=&   
\exp\left[\sum_i E^{(f_i)}(N_i)+E^{(f_i)}_{scale}(\mu_F,\mu_R)\right] \; 
\exp\left[ \sum_j {E'}^{(f_j)}(N_j)\right] 
\nonumber\\ && \hspace{-10mm} \times 
{\rm Tr} \left \{H\left(\alpha_s(\mu_R^2)\right) \; 
\bar{P} \exp \left[\int_{\sqrt{s}}^{{\sqrt{s}}/{\tilde N_j}} 
\frac{d\mu'}{\mu'} {\Gamma'}_S^\dagger\left(\alpha_s(\mu'^2)\right)\right] 
\right.
\nonumber\\ && \left. \times \,
{\tilde S} \left(\alpha_s(s/{\tilde N_j}^2) \right) \; 
P \exp \left[\int_{\sqrt{s}}^{{\sqrt{s}}/{\tilde N_j}} 
\frac{d\mu'}{\mu'}\; {\Gamma'}_S
\left(\alpha_s(\mu'^2)\right)\right] \right\} 
\nonumber 
\label{resHS}
\eeqa
with, in the $\overline{\rm MS}$ scheme, 
\beqa
\hspace{-10mm}
E^{(f_i)}(N_i)&=&
-\int^1_0 dz \frac{z^{N_i-1}-1}{1-z}\;
\left \{\int^{\mu_F^2}_{(1-z)^2s} \frac{d\mu'^2}{\mu'^2}
A^{(f_i)}\left(\alpha_s({\mu'}^2)\right) \right.
\nonumber \\ && \left. \hspace{35mm}
{}+{\nu}_{f_i}\left[\alpha_s((1-z)^2 s)\right]\right\} \, .
\eeqa
$E^{(f_i)}_{scale}$ describes the factorization and renormalization scale 
dependence of the cross section, while
${E'}^{(f_j)}$ appears if there are any massless final-state partons at 
lowest order.
$H$ and $S$ are the hard and soft functions and $\Gamma'_S$ the soft anomalous
dimensions, all matrices in color space in general (but they reduce to simple 
functions for processes with simple color flows) and known at lowest order
for most processes. For more details see Ref. \cite{NKNNLO}. 

Expanding the resummed cross section in $\alpha_s$ and inverting back from
moment space we can derive master formulas at fixed order for any
process.
The master formula for the next-to-leading-order (NLO) soft and virtual 
corrections for processes with simple color flow in 
the $\overline{\rm MS}$ scheme is
\beq
{\hat{\sigma}}^{(1)}
= \sigma^B \frac{\alpha_s(\mu_R^2)}{\pi}
\left\{c_3 \, {\cal D}_1(x_{th}) + c_2 \, {\cal D}_0(x_{th}) 
+c_1 \, \delta(x_{th})\right\} 
\label{NLOsimple}
\eeq
where $\sigma^B$ is the Born term, $c_3=\sum_i 2C_{f_i} -\sum_j C_{f_j}$,
\beqa
c_2&=&2 \, {\rm Re}{\Gamma'}_S^{(1)}- \sum_i \left[C_{f_i}
+2 C_{f_i} \, \delta_K \, \ln\left(\frac{-t_i}{M^2}\right)+
C_{f_i} \ln\left(\frac{\mu_F^2}{s}\right)\right]
\nonumber \\ && \hspace{-5mm}
{}-\sum_j \left[{B'}_j^{(1)}+C_{f_j}
+C_{f_j} \, \delta_K \, \ln\left(\frac{M^2}{s}\right)\right] 
\equiv T_2-\sum_i C_{f_i} \ln\left(\frac{\mu_F^2}{s}\right) 
\eeqa
and $c_1 =c_1^{\mu} +T_1$, with
\beq
c_1^{\mu}=\sum_i \left[C_{f_i}\, \delta_K \, \ln\left(\frac{-t_i}{M^2}\right)
-\gamma_i^{(1)}\right]\ln\left(\frac{\mu_F^2}{s}\right)
+d_{\alpha_s} \frac{\beta_0}{4} \ln\left(\frac{\mu_R^2}{s}\right) \, .
\eeq
We note that the $C_f$'s are color factors, $M$ is any appropriate hard scale, 
$\delta_K$ is 0 (1) for PIM (1PI) kinematics, and the sum over $j$ is only
relevant if we use 1PI kinematics and there are massless final-state partons
at lowest-order. More details and definitions are given in \cite{NKNNLO}
where results are also given in the DIS scheme.

The master formula for the NNLO soft and virtual corrections for 
processes with simple color flow 
in the $\overline{\rm MS}$ scheme is
${\hat\sigma}^{(2)}=
\sigma^B \, (\alpha_s^2(\mu_R^2)/\pi^2) \, {\hat{\sigma'}}^{(2)}$ with
\beqa
{\hat{\sigma'}}^{(2)}&=& 
\frac{1}{2} c_3^2 \, {\cal D}_3(x_{th})
+\left[\frac{3}{2} c_3 \, c_2 
- \frac{\beta_0}{4} c_3
+\sum_j C_{f_j} \frac{\beta_0}{8}\right] {\cal D}_2(x_{th})
\nonumber \\ && \hspace{-10mm}
{}+\left\{c_3 \, c_1 +c_2^2
-\zeta_2 \, c_3^2 -\frac{\beta_0}{2} \, T_2 
+\frac{\beta_0}{4} c_3  \ln\left(\frac{\mu_R^2}{s}\right)
+\sum_i C_{f_i} \, K \right.
\nonumber \\ && \hspace{-10mm} \quad  \left.
{}+\sum_j C_{f_j} \left[-\frac{K}{2} 
+\frac{\beta_0}{4} \, \delta_K  \ln\left(\frac{M^2}{s}\right)\right]
-\sum_j\frac{\beta_0}{4} {B'}_j^{(1)} \right\}
{\cal D}_1(x_{th})
\nonumber \\ && \hspace{-10mm} 
{}+\left\{c_2 \, c_1 -\zeta_2 \, c_2 \, c_3+\zeta_3 \, c_3^2 
-\frac{\beta_0}{2} T_1
+\frac{\beta_0}{4}\, c_2 \ln\left(\frac{\mu_R^2}{s}\right) 
+2 \, {\rm Re}{\Gamma'}_S^{(2)}-\sum_i {\nu}_{f_i}^{(2)}
\right. 
\nonumber \\ &&  \hspace{-10mm} \quad 
{}+\sum_i C_{f_i} \left[\frac{\beta_0}{8} 
\ln^2\left(\frac{\mu_F^2}{s}\right)
-\frac{K}{2}\ln\left(\frac{\mu_F^2}{s}\right)
-K \, \delta_K  \ln\left(\frac{-t_i}{M^2}\right)\right]
\nonumber \\ && \hspace{-10mm} \quad
{}-\sum_j \left({B'}_j^{(2)}+\nu_j^{(2)}\right)
+\sum_j C_{f_j} \, \delta_K \left[\frac{\beta_0}{8}
\ln^2\left(\frac{M^2}{s}\right)
-\frac{K}{2}\ln\left(\frac{M^2}{s}\right)\right]
\nonumber \\ && \hspace{-10mm} \quad \left.
{}-\sum_j \frac{\beta_0}{4}
{B'}_j^{(1)} \delta_K \ln\left(\frac{M^2}{s}\right) \right\}
{\cal D}_0(x_{th})  
+R_{\delta(x_{th})} \, \delta(x_{th}) \, .
\label{NNLO}
\eeqa
For more details and explicit expressions for the full $\delta(x_{th})$ 
terms see Ref. \cite{NKNNLO}.

For the more general case of complex color flow, where
$H$, $S$, and $\Gamma'_S$ are matrices,  the $\overline{\rm MS}$ scheme
NLO master formula generalizes to 
\beq
{\hat{\sigma}}^{(1)} = {\hat{\sigma}}^{(1)}_{\rm simple}
+\frac{\alpha_s^{d_{\alpha_s}+1}(\mu_R^2)}{\pi} 
\left[A^c \, {\cal D}_0(x_{th})+T_1^c \, \delta(x_{th})\right] \, ,
\label{NLOcx}
\eeq
where  ${\hat{\sigma}}^{(1)}_{\rm simple}$ denotes the 
result in Eq. (\ref{NLOsimple}) but without the 
$2{\rm Re}{\Gamma'}_S^{(1)}$ term in $c_2$, $d_{\alpha_s}$ denotes
the power of $\alpha_s$ in the Born term, and
\beq
A^c={\rm Tr} \left(H^{(0)} {\Gamma'}_S^{(1)\,\dagger} S^{(0)}
+H^{(0)} S^{(0)} {\Gamma'}_S^{(1)}\right) \, . 
\eeq

At NNLO for processes with complex color flow
the $\overline{\rm MS}$ scheme master formula generalizes to 
\beqa
{\hat{\sigma}}^{(2)}&=&{\hat{\sigma}}^{(2)}_{\rm simple}
+\frac{\alpha_s^{d_{\alpha_s}+2}(\mu_R^2)}{\pi^2} 
\left\{\frac{3}{2} c_3 \, A^c\,
{\cal D}_2(x_{th}) \right.
\nonumber \\ && 
{}+\left[\left(2c_2-\frac{\beta_0}{2}\right)A^c 
+c_3 \, T_1^c+F\right] {\cal D}_1(x_{th})
\nonumber \\ &&
{}+\left[\left(c_1-\zeta_2 c_3+\frac{\beta_0}{4}
\ln\left(\frac{\mu_R^2}{s}\right)\right) A^c 
+\left(c_2-\frac{\beta_0}{2}\right) T_1^c \right.
\nonumber \\ && \quad \left. \left.
{}+F \, \delta_K \ln\left(\frac{M^2}{s}\right)+G\right] {\cal D}_0(x_{th})
+R^c_{\delta(x_{th})} \, \delta(x_{th}) \right\} ,
\label{NNLOc}
\eeqa
with ${\hat{\sigma}}^{(2)}_{\rm simple}$ denoting Eq. (\ref{NNLO})
times $\alpha_s^2(\mu_R^2)/\pi^2$ (but without any 
${\Gamma'}_S$ terms), and 
\beq
F={\rm Tr} \left[H^{(0)} \left({\Gamma'}_S^{(1)\,\dagger}\right)^2 S^{(0)}
+H^{(0)} S^{(0)} \left({\Gamma'}_S^{(1)}\right)^2
+2 H^{(0)} {\Gamma'}_S^{(1)\,\dagger} S^{(0)} {\Gamma'}_S^{(1)} \right] .
\eeq
Again, for more details and the full explicit $\delta(x_{th})$ terms 
see Ref. \cite{NKNNLO}.
Eq. (\ref{NNLOc}) serves as the most general master formula for the 
NNLO soft and virtual corrections for any process in hadron-hadron
and lepton-hadron collisions.

\section{Applications to QCD processes}
Using the NNLO master formula I have rederived known NNLO results
for Drell-Yan and related processes and for deep inelastic-scattering (DIS),
and produced new results for many other processes \cite{NKNNLO}. 
Here I give a few examples.

For deep inelastic scattering, $\gamma^* q \rightarrow q$, 
in the $\overline{\rm MS}$ scheme, we have
$c_3=C_F$, $c_2=-3C_F/4 - C_F \ln(\mu_F^2/Q^2)$, and
$c_1=(-3/4) C_F \ln(\mu_F^2/Q^2)-C_F \zeta_2-(9/4) C_F$.
The NNLO corrections from the master formula then agree with the
results for the coefficient functions in Ref. \cite{DIS}, and we
also derive the two-loop function ${B'}_q^{(2)}$ which is needed
in NNLL resummation \cite{NKNNLO}.
In the DIS scheme, the NLO and NNLO corrections vanish, as expected.

For heavy quark hadroproduction in the channel
$q {\bar q} \rightarrow Q {\bar Q}$ in 1PI kinematics, we have
$c_3=4C_F$, $c_2=-2C_F-2C_F\ln(t_1u_1/m^4)-2C_F\ln(\mu_F^2/s)$,
$c_1^{\mu}=C_F [\ln(t_1u_1/m^4)-3/2]\ln(\mu_F^2/s) 
+(\beta_0/2)\ln(\mu_R^2/s)$,
$A^c=(\sigma^B/ \alpha_s^2) \, 2\, {\rm Re}{\Gamma'}_{S,22}^{(1)}$.
The NNLO corrections are in agreement with and extend earlier
NNLO-NNLL calculations \cite{KLMV}.
Results have also been obtained in PIM kinematics and in the DIS scheme,
and also for the $gg \rightarrow Q {\bar Q}$ channel.

A final application is jet production, where many processes 
are involved. For example, for $g g \rightarrow  g g$ we have
$c_3=2C_A$ and $c_2=-2C_A\ln(\mu_F^2/s)-\beta_0/2-4C_A$.
The NNLO corrections extend earlier NNLO-NLL calculations \cite{KO2}.


\begin{thebibliography}{0}

\bibitem{NKNNLO}
N. Kidonakis, Cavendish-HEP-03/02, hep-ph/0303186.

\bibitem{KS}
N. Kidonakis and G. Sterman, Phys. Lett. B {\bf 387}, 867 (1996);
Nucl. Phys. {\bf B505}, 321 (1997); N. Kidonakis, Int. J. Mod. Phys. 
A {\bf 15}, 1245 (2000).

\bibitem{KOS}
N. Kidonakis, G. Oderda, and G. Sterman,
Nucl. Phys. {\bf B525}, 299 (1998);
Nucl. Phys. {\bf B531}, 365 (1998).

\bibitem{LOS}
E. Laenen, G. Oderda, and G. Sterman, 
Phys. Lett. B {\bf 438}, 173 (1998).

\bibitem{KDO}
N. Kidonakis and V. Del Duca, Phys. Lett. B {\bf 480}, 87 (2000);
N. Kidonakis and J.F. Owens, Phys. Rev. D {\bf 61}, 094004 (2000);
N. Kidonakis, Nucl. Phys. B (Proc. Suppl.) {\bf 79}, 410 (1999).

\bibitem{DIS}
E.B. Zijlstra and W.L. van Neerven,
Nucl. Phys. {\bf B383}, 525 (1992).

\bibitem{KLMV} N. Kidonakis, Phys. Rev. D {\bf 64}, 014009 (2001); 
N. Kidonakis, E. Laenen, S. Moch, and R. Vogt, 
Phys. Rev. D {\bf 64}, 114001 (2001); D {\bf 67}, 074037 (2003).  

\bibitem{KO2}
N. Kidonakis and J.F. Owens, Phys. Rev. D {\bf 63}, 054019 (2001).

\end{thebibliography}
\end{document}